\documentclass[twocolumn,preprintnumbers,amsmath,amssymb]{revtex4-2}
\usepackage{graphicx}
\usepackage[dvipsnames]{xcolor}
\usepackage{ulem}
\usepackage{url,hyperref}
\usepackage{lipsum}
\newcommand{\be}{\begin{equation}}
\newcommand{\ee}{\end{equation}}
\newcommand{\beq}{\begin{equation}}
\newcommand{\eeq}{\end{equation}}
\newcommand{\bea}{\begin{eqnarray}}
\newcommand{\eea}{\end{eqnarray}}
\newcommand{\eml}{\end{mathletters}}

\newcommand{\hf}{|\text{HF}\rangle}
\newcommand{\bcs}{|\text{BCS}\rangle}
\newcommand{\pbcs}{|\text{PBCS}\rangle}
\newcommand{\sccd}{|\text{vCCD}_{\text{sep}}\rangle}
\newcommand{\ccd}{|\text{vCCD}\rangle}

\begin{document}

\title{Variational theory combining number-projected BCS and coupled-cluster doubles}
\author{V.V. Baran$^{1,2}$ }
\email[]{vvbaran@fizica.unibuc.ro}
\author{J. Dukelsky$^{3}$}
\email[]{j.dukelsky@csic.es}
\affiliation{
$^1$ Faculty of Physics, University of Bucharest,
405 Atomi\c stilor, POB MG-11, Bucharest-M\u agurele, RO-077125, Romania\\
$^2$"Horia Hulubei" National Institute of Physics and
Nuclear Engineering, \\
30 Reactorului, RO-077125, Bucharest-M\u agurele, Romania \\
$^3$ Instituto de Estructura de la Materia, CSIC, Serrano 123, E-28006 Madrid, Spain
}

\begin{abstract}
The ground state pairing correlations in finite fermionic systems are described with a high degree of accuracy within a variational approach based on a combined coupled-cluster and particle-number-projected BCS ansatz. The flexibility of this symmetry-preserving wavefunction enables a unified picture valid from weak to strong coupling, both in small and large systems. The present variational approach consistently yields an energy upper bound while operating at the same level of precision of the non-variational particle-number projected Bogoliubov-coupled-cluster theory [Phys. Rev. C 99, 044301 (2019)].

\end{abstract}

\maketitle

\section{Introduction}
Pairing Hamiltonians are ubiquitous in quantum many-body physics. Starting from their variational treatment in the microscopic theory of superconductivity given by Bardeen, Cooper and Schrieffer (BCS) \cite{bcs}, they were soon exported to nuclear physics for the descripton of the large gaps observed in even-even nuclei \cite{Bohr}. However, the violation of particle number in the BCS theory, which is negligible for macroscopic systems, represents a major drawback when applied to finite systems. 
Therefore, techniques to implement number projection on top of the BCS wavefunction (PBCS) were developed in nuclear structure \cite{mang64, Sheik2000} and more recently in quantum chemistry \cite{Scus2011, Kamoshi2019} where the PBCS wavefunction is known as the antisymmetrized geminal power (AGP). PBCS improves over the BCS theory in finite systems, specially in the strong coupling limit where superconductivity is well established, but it still fails in the weak coupling limit dominated by pairing fluctuations, and all along the transitional region \cite{Sandu2008, duk16}. This fact was made evident in ultrasmall superconducting grains, where PBCS predicted an abrupt metal-superconductor transition as a function of the grain size \cite{Delft1998} while the exact solution showed a smooth crossover dominated by large fluctuations \cite{Duke1999}. It was precisely in the field of ultrasmall superconducting grains that the exact solution of the constant pairing Hamiltonian given by Richardson in the sixties \cite{richardson63} was recovered \cite{Sierra2000} and intensively used as a natural benchmark model for superconducting theories beyond BCS \cite{HIRSCH2002, DUKE2003, henderson2014, degroote16, Rubio2018, qiu2019,  dutta2021, doi:10.1063/1.5116715, doi:10.1063/5.0021144}. 

In the extreme weak coupling limit pair coupled-cluster doubles (pCCD) describes correctly the pairing fluctuating regime but it quickly overbinds due to the non-variational character of the theory based on the left projection to a Hartree-Fock Slater determinant \cite{duk2003, Henderson2015, duk16}. It seems, therefore, that a combination of pCCD and quasiparticle BCS would be able to approach both limits correctly. Indeed, the extension of pCCD to BCS quasiparticles, the BCS-CCD method \cite{henderson2014}, gave the correct behavior in the weak coupling limit, but still suffers from large deviations ($\thicksim$ 10 \%  for sizes of $\thicksim$ 100 particles) across the transition region. There were attempts to interpolate between pCCD and PBCS \cite{degroote16, duk16} or to diagonalize the pairing Hamiltonian in a subspace defined by the reference PBCS state and different two and four number projected quasiparticle states \cite{PhysRevC.95.014326}. Other possible ways to add correlations to the PBCS state were explored in  \cite{Henderson_2020, Kamoshi2021}.

Perhaps, the most successful theory beyond BCS-CCD is simply its number projected version coined as particle-number projected BCS coupled-cluster doubles (PBCS-CCD) theory \cite{qiu2017}. The theory is not Ritz variational and therefore, it cannot assure an upper bound for the ground state energy. However it has an affordable computational cost that scales polynomially with the system size. Moreover, it gives excellent numerical results both in the weak and strong coupling limits, as well as in the transitional region. 

The aim of our paper is to design a Ritz variational method with a trial wavefunction that combines pCCD and PBCS and produces numerical results with the same level of precision of PBCS-CCD and with a similar computational cost. We will benchmark our variational theory with the exact solution of the Richardson model \cite{richardson66} and with PBCS-CCD results \cite{qiu2017} where available.

\section{Theoretical background}
\subsection{Pairing Hamiltonian}

We consider the generic pairing Hamiltonian
\beq
\label{ham}
H=\sum_{i=1}^L \epsilon_i\, (c^\dagger_i c_i + c^\dagger_{\bar{i}} c_{\bar{i}}) + \sum_{i,j=1}^L V_{i,j} c^\dagger_{i}c^\dagger_{\bar{i}} c_{\bar{j}} c_j~,
\eeq
where $i$ and $\bar{i}$ indicate one of the $L$ pairs  of conjugated degenerate single particle levels with energy $\epsilon_i=\epsilon_{\bar{i}}$. 

The Hamiltonian (\ref{ham}) preserves seniority, and for simplicity we restrict ourselves to the seniority zero ($v=0$) subspace . Then, in the absence of the interaction term the ground state is given by the Hartree-Fock product state
\beq
\label{hf}
\hf=\prod_{i=1}^M c^\dagger_{i}c^\dagger_{\bar{i}} \,|0\rangle~,
\eeq
where $M$ is the number of pairs in the system.

In preparation for the discussion that follows, we pass to the particle-hole (ph) representation and denote the particle levels with $p>M$ and the hole levels with $h\leq M$. We define the particle and hole pair and number operators
\beq
\label{pairops}
\begin{aligned}
P^\dagger_p&=c^\dagger_p c^\dagger_{\bar{p}}~,~ P_p=c_{\bar{p}}c_{p}~,~N_{p}=c^\dagger_p c_p + c^\dagger_{\bar{p}} c_{\bar{p}}~,\\
P^\dagger_h&=c_{\bar{h}}c_{h}~,~ P_h=c^\dagger_h c^\dagger_{\bar{h}}~,~N_{h}=2-c^\dagger_h c_h - c^\dagger_{\bar{h}} c_{\bar{h}}~,
\end{aligned}
\eeq
such that the Hartree-Fock state (\ref{hf}) is the vacuum to the $P$ and $N$ operators, $P_p\hf=P_h\hf=0$, $N_p\hf=N_h\hf=0$. The pairing Hamiltonian (\ref{ham}) is then expressed as
\begin{equation}
\label{hamph}
    \begin{aligned}
    H =&~ E_{\text{HF}}+\sum_{p=M+1}^L \epsilon_p\, N_p+ \sum_{h=1}^M (-\epsilon_h-V_{hh})\, {N}_h\\
         &+\sum_{p,p'=M+1}^L V_{pp'}P^\dagger_p P_{p'}+\sum_{h,h'=1}^M V_{hh'}{P}^\dagger_h {P}_{h'}\\
         &+\sum_{p=M+1}^L\sum_{h=1}^M V_{ph}\left(P^\dagger_p {P}^\dagger_{h}+{P}_h P_{p}\right)~,
    \end{aligned}
\end{equation}
with $E_{\text{HF}}=\langle \text{HF} | H |\text{HF}\rangle=\sum_{h=1}^M\left(2\epsilon_h+V_{h h}\right)$ being the energy of the Hartree-Fock state (\ref{hf}).

\subsection{Mean-field theory and its symmetry restoration}

The standard description of the pairing correlations induced by the Hamiltonian (\ref{ham}) is given within the BCS approximation \cite{bcs} in terms of the pair condensate
\beq
\bcs=\exp[\Gamma^\dagger(x)]|0\rangle, \quad \Gamma^\dagger(x)\equiv\sum_{i=1}^L x_i c^\dagger_i c^\dagger_{\bar{i}}~,
\eeq
which explicitly breaks the $U(1)$ gauge symmetry associated with particle number conservation. For macroscopic systems the symmetry broken picture is exact and the particle number fluctuations are negligible. However, for finite systems like atomic nuclei or small superconducting grains one speaks only of \textit{obscured} or \textit{emergent}  symmetry  breaking \cite{ui83,koma94,Yannouleas_2007,papenbrock2014}, in which case the quantum fluctuations inevitably lift any degeneracy associated with the broken symmetry.

Much effort is thus devoted to restore the symmetry of the mean-field ansatz with the help of projection techniques \cite{lowdin55,mayer80,bender03,SCHMID2004,jimenez2012,RS,BR}. In the BCS case upon  particle number restoration we obtain the so-called number-projected BCS (PBCS) \cite{mang64} or antisymmetrized geminal power (AGP) in the context of quantum chemistry \cite{coleman65}
\beq
\label{pbcs}
\begin{aligned}
|\text{PBCS}(x)\rangle&=\mathcal{P}_M\bcs\\
&=\frac{1}{M!}[\Gamma^\dagger(x)]^M|0\rangle\\
&= \frac{1}{2\pi}\int_0^{2\pi} \text{d}\theta ~e^{-i\theta M }\exp[\Gamma^\dagger(e^{i\theta}x)]~,
\end{aligned}
\eeq
where $\mathcal{P}_M$ is the projector onto the state of $M$ pairs. While PBCS describes well the properties of superfluid nuclei with a small number of valence nucleons \cite{SCHMID2004},
it cannot account for the weak pairing correlations that develop within larger spaces, e.g. those considered in the large-scale
energy density functional treatments of finite nuclei or in the study of small superconducting grains \cite{duk99}. For a working description of the weak pairing regime one usually turns to RPA \cite{duk2003} or coupled-cluster \cite{henderson2014, Henderson2015, duk16} approaches. Generalizations of the PBCS ansatz have also been considered based on its structural similarity with a particular coupled-cluster ansatz \cite{duk16}. Specifically, the PBCS representation in the ph-basis is obtained as
\beq
\label{phpbcs}
\begin{aligned}
	&|\mathrm{PBCS}\rangle\propto\sum_{\ell=0}^{M} \frac{1}{\ell !^{2}}[\Gamma_{P}^{\dagger}(x) \Gamma^\dagger_{H}(1 / x)]^{\ell}\, \hf\\
	& =\frac{1}{2\pi}\int_0^{2\pi} \text{d}\theta~\exp[\Gamma^\dagger_P(e^{i\theta}x)]~ \exp[\Gamma^\dagger_H(e^{-i\theta}/x)]\, |\text{HF}\rangle~,
\end{aligned}
\eeq
in terms of the particle and hole components of the collective pairs
\beq
\label{gammaph}
\Gamma_{P}^{\dagger}(x)=\sum_{p=M+1}^{L} x_{p} P_{p}^{\dagger}, \quad \Gamma_{H}^{\dagger}(x)=\sum_{h=1}^{M} x_{h} P_{h}^{\dagger}~.
\eeq
The structure of the PBCS state is then defined by the inverse squared factorials appearing as expansion coefficients in the collective ph-pair basis.

\subsection{Coupled-cluster theory}

Analogously, by using a slightly modified expansion involving plain factorials one obtains a separable pair coupled-cluster doubles variational ansatz (vCCD$_{\text{sep}}$)
\beq
\label{sccd}
\begin{aligned}
\sccd&=\sum_{\ell=0}^{M} \frac{1}{\ell !}[\Gamma_{P}^{\dagger}(y) \Gamma^\dagger_{H}(y)]^{\ell}\, \hf\\
&=\exp[\Gamma_{P}^{\dagger}(y) \Gamma^\dagger_{H}(y)]\,\hf\\
&=\exp\left[\sum_{p,h}y_p y_h P^\dagger_p P^\dagger_h\right]\,\hf~.
\end{aligned}
\eeq
This is a particular case of a pair coupled-cluster doubles variational  wavefunction (vCCD)
\beq
\label{ccd}
\ccd=\exp\left[\sum_{p,h}z_{ph} P^\dagger_p P^\dagger_h\right]\,\hf~,
\eeq
involving the most general double excitations that do not break
pairs through a fully non-separable structure matrix $z_{ph}$ \cite{duk2003}. The separable case vCCD$_{\text{sep}}$ is thus recovered for $z_{ph}=y_p\, y_h$.

On the one hand, the full freedom in the structure matrix of the CCD ansatz allows for an excellent description of the weak pairing regime (see also Fig. \ref{fig1} below), due to the additional four-body correlations accounted for relative to the separable case. Note, for example, that the Richardson solution involving complex-conjugated pairs may be expressed exactly as a product of four-body quartet structures \cite{sam13}. On the other hand, the computational complexity of evaluating operator matrix elements exactly in the non-separable case grows exponentially with the size of the system. The usual approximation of coupled cluster involves a left projection onto the subspaces of zero and two ph pairs, thus breaking the Ritz variational principle. Furthermore, even with a fully nonseparable structure matrix, the validity of the vCCD ansatz breaks down around the critical value of the pairing strenght (see also Fig. \ref{fig1} below). The choice of a BCS mean-field reference state does improve on this aspect at the cost of effectively breaking the particle number symmetry \cite{henderson2014}.

Given the success of vCCD in the weak pairing regime and that of PBCS in the strong pairing regime, it is then natural to combine them for a precise unified description of all regimes.

\subsection{Combining coupled-cluster and symmetry-restored mean-field theories}

The symmetry restoration of broken-symmetry coupled-cluster theories has been only recently considered \cite{duguet2015,duguet2017,qiu2017,tsuchimochi2017}. For the schematic pairing Hamiltonian with the breaking and restoration of the particle number symmetry, the so-called particle-number projected Bogoliubov-coupled-cluster theory yields highly accurate results \cite{qiu2019}. In this context, a set of differential equations is set up for obtaining the  gauge-angle-dependent excitation operator. The practical need for truncating this set of ODEs implies however an approximate action of the projection operator and a violation of the Ritz variational principle.

It is the purpose of this work to explore a physically transparent alternative in the form of a variational CCD-PBCS combined approach. Ideally the ground state would involve the CCD excitations built directly on top of the particle-number-projected BCS as
\beq
\label{ccdpbcs}
|\text{vCCD-PBCS}\rangle=\exp\left[\sum_{p,h}z_{ph} P^\dagger_p P^\dagger_h\right]|\text{PBCS}(x)\rangle~,
\eeq
where the $L$ mixing amplitudes $x_i$ of the PBCS state and the $M(L-M)$ structure matrix elements $z_{ph}$ of the CCD excitations are to be treated as free variational parameters. Indeed, the energy minimization procedure
\beq
\label{energy}
E_\text{gs}=\text{min}_\psi\frac{\langle \psi | H |\psi\rangle}{\langle \psi| \psi \rangle }
\eeq
yields extremely precise results for the few small systems where this ansatz may actually be applied.

Computational access to larger systems is enabled upon various simplifications.  Let us consider first the separability approximation $z_{ph}=y_p y_h$ for the structure matrix of the CCD excitations in Eq. (\ref{ccdpbcs}), yielding the $\text{vCCD}_{\text{sep}}\text{PBCS}$ ansatz
\beq
\label{sccdpbcs}
\begin{aligned}
|\text{vCCD}_{\text{sep}}\text{PBCS}\rangle&=\exp[\Gamma_{P}^{\dagger}(y) \Gamma^\dagger_{H}(y)]\,|\text{PBCS}(x)\rangle~,
\end{aligned}
\eeq
which involves $2L$ independent variational parameters $x_i, y_i$.
Remarkably, while the PBCS ansatz (\ref{pbcs}) and the $\text{vCCD}_{\text{sep}}$ ansatz (\ref{sccd}) each individually fails in the weak pairing regime due to their separable structure matrices, the combined $\text{vCCD}_{\text{sep}}\text{PBCS}$ wavefunction will turn out to be quite accurate. This may be easily understood by noting that for weak pairing the above ansatz effectively involves a non-separable structure matrix in the form
\beq
\label{sccdpbcs_weak}
|\text{vCCD}_{\text{sep}}\text{PBCS}\rangle\approx [1+\sum_{p,h}(x_px_h+y_py_h)P^\dagger_p P^\dagger_h]\hf,
\eeq
obtained after taking into account the PBCS (\ref{pbcs}) and $\text{vCCD}_{\text{sep}}$ (\ref{sccd}) expansions and redefining $x_h\rightarrow 1/x_h$ in (\ref{phpbcs}). The accuracy of this doubly-separable $2L$-dimensional parametrization of the fully non-separable $M(L-M)$-dimensional structure matrix naturally degrades with increasing system size, the actual rate being numerically determined in the next section.

An alternative approximation scheme would then involve a fully non-separable CCD excitations limited to a relatively small finite window around the Fermi level. We thus consider the combination
\beq
\label{ccdwpbcs}
\begin{aligned}
&|\text{vCCD}^{(w)}\text{PBCS}(x,z)\rangle=\text{vCCD}^{(w)}(z)|\text{PBCS}(x)\rangle
\end{aligned}
\eeq
with
\beq
\label{ccdwpbcs2}
\begin{aligned}
&\text{vCCD}^{(w)}(z)=\exp\left[\sum_{p=M+1}^{M+w/2}\sum_{h=M+1-w/2}^Mz_{ph} P^\dagger_p P^\dagger_h\right]~,
\end{aligned}
\eeq
where $w$ denotes the size of the truncation window. As will be detailed in the next section, the quality of the results obtained within this approach will turn out to be inferior to that of the above $\text{vCCD}_{\text{sep}}\text{PBCS}$ ansatz of Eq. (\ref{sccdpbcs}) involving two sets of global parameters.

Overall the optimal compromise between the computational complexity and the accuracy of the results is found by combining the above two approximation schemes (\ref{sccd}) and (\ref{ccdwpbcs}) into the ansatz
\beq
\label{ccdwseppbcs}
\begin{aligned}
&|\text{vCCD}^{(w)}_{\text{sep}}\text{PBCS}(x,y,z)\rangle=\text{vCCD}^{(w)}_{\text{sep}}(y,z)|\text{PBCS}(x)\rangle~,\\
&\text{vCCD}^{(w)}_{\text{sep}}=\text{vCCD}^{(w)}(z)\, \text{vCCD}_{\text{sep}}(y)~,
\end{aligned}
\eeq
involving a total of $2L+(w/2)^2$ free variational parameters and leading to highly accurate results comparable to those of Ref. \cite{qiu2019}, to be discussed later on. Next, we shortly review the actual computational strategy for the above mentioned wavefunctions.

\subsection{Computational aspects}

In this section we propose a novel efficient algorithm for the evaluation of expectation values on the combined $\text{vCCD}^{(w)}_{\text{sep}}\text{PBCS}$ wavefunction (\ref{ccdwseppbcs}). It is based on  a representation of the PBCS (\ref{phpbcs}), vCCD$_{\text{sep}}$ (\ref{sccd_op}) and vCCD$^{(w)}$ (\ref{ccdwpbcs2}) terms as disentangled particle and hole gauge-angle-rotated BCS states.

We start from the discrete exact representation for the particle-number projection operation and decompose the PBCS ansatz using the definitions of Eqs. {(\ref{phpbcs}) and} (\ref{gammaph})
\beq
\label{pbcs_expans}
\begin{aligned}
\pbcs =\sum_{n=0}^{L} \exp[\Gamma^\dagger_H(e^{-i\theta_n}x)]\exp[\Gamma^\dagger_P(e^{i\theta_n}x)]\, |\text{HF}\rangle~,
\end{aligned}
\eeq
where $\theta_n=2\pi n/(L+1)$. Note that we neglect the irrelevant constant normalization factors throughout this section.

The collective pairs (for the  particle and hole subspaces) appearing in the vCCD$_{\text{sep}}$ operator
\beq
\label{sccd_op}
\text{vCCD}_{\text{sep}}=\sum_{\ell=0}^{M} \frac{1}{\ell !}[\Gamma_{P}^{\dagger}(y) \Gamma^\dagger_{H}(y)]^{\ell}
\eeq
are also expanded as superpositions of gauge-angle-rotated BCS operators
\begin{equation}
\label{Gamma_ccd_expans}
\begin{aligned}
\frac{[\Gamma_P^\dagger(y)]^\ell}{\ell!}&=\sum_{k=0}^{L-M}e^{-i\phi_k \ell}\exp[\Gamma_P^\dagger(e^{i\phi_k}y)]~,\\
\frac{[\Gamma_H^\dagger(y)]^\ell}{\ell!}&=\sum_{k=0}^M e^{-i\varphi_k \ell }\exp[\Gamma_H^\dagger(e^{i\varphi_k}y)]~,
             \end{aligned}
         \end{equation}
with $\phi_k=2\pi k/(L-M+1)$ and $\varphi_k=2\pi k/(M+1)$. Finally we obtain the representation
         \begin{equation}
         \label{sccdpbcs2}
             \begin{aligned}
|\text{vCCD}&_{\text{sep}}\text{PBCS}\rangle=\sum_{k_p=0}^{L-M}\sum_{k_h=0}^M\sum_{n=0}^L g_{k_pk_h}\,  \times \\
& \text{BCS}_P(k_p,n)\,\text{BCS}_H(k_h,n)\, \hf
             \end{aligned}
         \end{equation}
in terms of the gauge-angle-rotated BCS operators
\beq
\label{rotated_bcs}
\begin{aligned}
\text{BCS}_P(k_p,n)&=\exp[\Gamma_P^\dagger(e^{i\theta_n}x+e^{i\phi_{k_p}}y)]~,\\
\text{BCS}_H(k_h,n)&=\exp[\Gamma_H^\dagger(e^{-i\theta_n}x+e^{i\phi_{k_p}}y)]~,
\end{aligned}
\eeq
with expansion coefficients
\beq
\label{expans_coeffs_sccd}
g_{k_pk_h}=\sum_{\ell=0}^M \ell!\,\exp[-i(\phi_{k_p}+\varphi_{k_h})\ell]~.
\eeq

We are interested in the expectation values of generic particle-number-conserving ph-factorized operators $\mathcal{O}=\mathcal{O}^{(P)}\mathcal{O}^{(H)}$ on the $\text{vCCD}_{\text{sep}}\text{PBCS}$ state (\ref{sccdpbcs2}). As the particle-number projection operation needs only to be performed once on the mean-field BCS wavefunction, we obtain
\begin{equation}
\label{averages}
\begin{aligned}
&\langle\mathcal{O}\rangle=\,\langle \text{vCCD}_{\text{sep}}\text{BCS}|\mathcal{O}^{(P)}\mathcal{O}^{(H)}|\text{vCCD}_{\text{sep}}\text{PBCS}\rangle\\
&=\sum_{k_p,k_p'=0}^{L-M}\sum_{k_h,k_h'=0}^M g^*_{k_pk_h}\, g_{k_p'k_h'}\sum_{n=0}^L {O}^{(P)}_{k_p,k_p',n}\,\mathcal{O}^{(H)}_{k_h,k_h',n}~,
\end{aligned}
\end{equation}
involving the matrix elements between particle and hole BCS states (\ref{rotated_bcs})
\begin{equation}
\label{bcs_matrix_elem}
\begin{aligned}
{O}^{(P)}_{k_p,k_p',n}&=\langle \text{BCS}_P(k_p,0)|{O}^{(P)}|\text{BCS}_P(k_p',n)\rangle~,\\
\mathcal{O}^{(H)}_{k_h,k_h',n}&=\langle \text{BCS}_H(k_h,0)|{O}^{(H)}|\text{BCS}_H(k_h',n)\rangle~.
\end{aligned}
\end{equation}

By considering all terms in Eq. (\ref{hamph}), the expression (\ref{averages}) may be directly employed to compute the energy expectation value  $E=\langle H\rangle/\langle I\rangle$ on the vCCD$_{{\text{sep}}}$PBCS wavefunction.

To include the effects of the non-separable vCCD$^{(w)}$ excitations of Eq. (\ref{ccdwseppbcs}) we use a Hubbard-Stratonovich transformation \cite{quartets_plb} and pass to a disentangled BCS representation of the vCCD operator
\beq
\begin{aligned}
&\text{vCCD}(z)=\exp\left(\sum_{p,h}z_{ph} P^\dagger_p P^\dagger_h\right)\\
&=\int \text{d}^{L}\xi\, \exp\left(-\frac{1}{2}\, {\xi}^{T} \,Z^{-1} \,{\xi}\right)\,\exp[\Gamma^\dagger_P(\xi)+\Gamma^\dagger_H(\xi)],
\end{aligned}
\eeq
with
\begin{equation}
Z=\begin{pmatrix}
0 & z\\
z^T & 0
\end{pmatrix}~.
\end{equation}
This BCS representation allows us to generalize the above Eq. (\ref{averages}) to the expectation values of ph-factorized operators on the  vCCD$^{(w)}_{\text{sep}}$PBCS wavefunction (\ref{ccdwseppbcs}).
In practice the matrix elements on the resulting BCS wavefunctions which generalize Eq. (\ref{bcs_matrix_elem}) are to be treated as polynomials in the integration variables $\xi$. Only specific terms corresponding to the various nonzero Wick contractions are to be selected according to
\begin{equation}
\begin{aligned}
     \int \text{d}^{L}\xi&\, \exp\left(-\frac{1}{2}\, \xi^{T} \,Z^{-1} \,\xi\right)\, \xi_{i} \xi_{j} \dots \xi_{k} \xi_{l}=\\
     &\sum_{\text {Wick }} Z_{a b} \dots Z_{c d}~,
\end{aligned}
\end{equation}
the set of indices $\{a, b, . . . , c, d\}$ representing a permutation of $\{i , j , . . . , k, l\}$ \cite{zee}.

With the non-separable CCD excitations restricted to a small window of $w$ levels around the Fermi level as in Eq. (\ref{ccdwpbcs}), the matrix elements of the relevant operators on generic vCCD$^{(w)}$BCS states may be computed analytically (the notebook in Cadabra2 \cite{cadabra} is available {upon request from the authors}). Their subsequent coupling to the separable subspace (i.e. the complement of the $w$-level window) leads to the final   vCCD$^{(w)}_{\text{sep}}$PBCS results presented in the next section.

\section{Numerical results}
In this section we benchmark the variational calculations for the various wavefunctions presented above against the exact solution \cite{richardson66} for a constant pairing Hamiltonian ($V_{i,j}=-G$ in Eq. (\ref{ham})) of equally spaced single particle levels with energies $\epsilon_k=k \epsilon,\, k=1,\dots,L$. In order to avoid the singularities in the equations that solve exactly the picket fence model we use the same algorithm proposed by Richardson in \cite{richardson66}.

The  numerical code used to compute the expectation value of the Hamiltonian (\ref{hamph}) on the various derivatives of the vCCD-PBCS wavefunction is freely available {upon request from the authors}. The minimization procedure for the energy function (\ref{energy}) is performed using the e04ucf routine of the NAG library \cite{nag}.

\begin{figure}[t]
\centering
\includegraphics[width=0.5\textwidth]{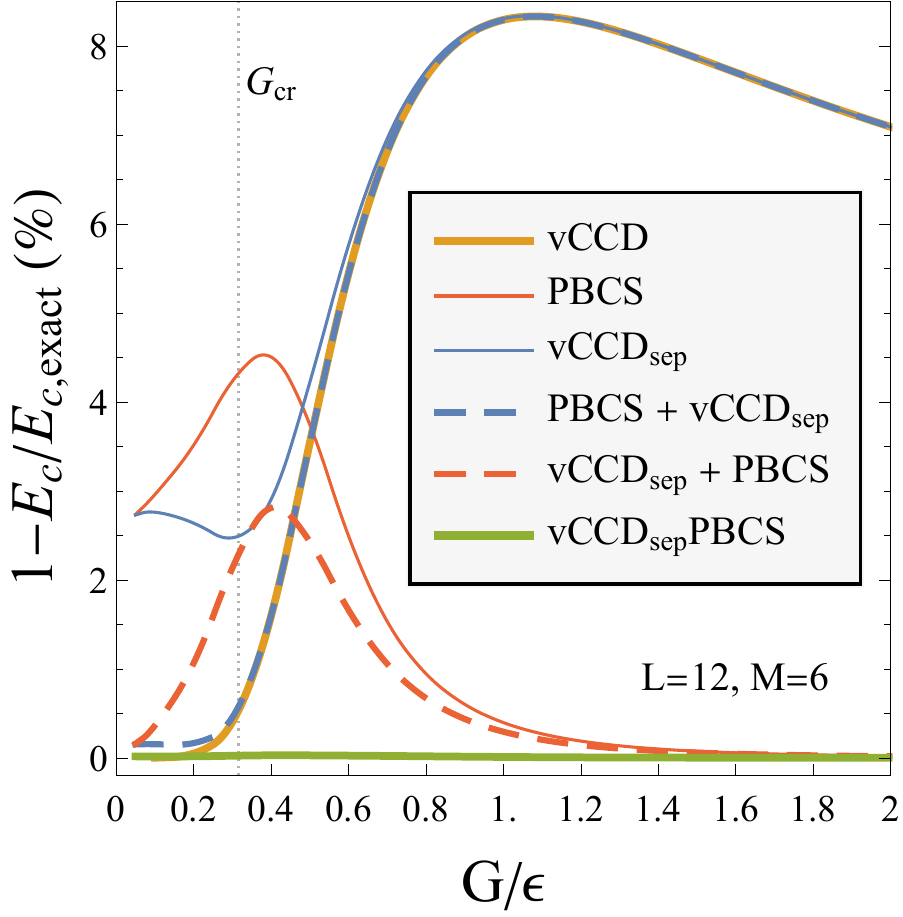}
\caption{Error in the correlation energy  (\ref{corr_energy}) relative to its exact value (in percentages) versus the pairing strength $G$ (in units of the level spacing $\epsilon$) for the fully variational vCCD (\ref{ccd}), PBCS (\ref{pbcs}), and vCCD$_{\text{sep}}$PBCS (\ref{sccdpbcs}) wavefunctions in the case of an $L=12$ level system at half filling. The ``vCCD$_{\text{sep}}$+PBCS'' procedure indicated by a red dashed line involves an energy minimization with respect to the vCCD$_{\text{sep}}$ amplitudes in the presence of the fixed optimal PBCS reference. Within the ``PBCS+vCCD$_{\text{sep}}$''  procedure indicated by a green dashed line the PBCS amplitudes are varied in the presence of the fixed optimal vCCD$_{\text{sep}}$ structure.}
\label{fig1}
\end{figure}

We present in Figure (\ref{fig1}) the errors for the correlation energy
\beq
\label{corr_energy}
E_c=\frac{\langle \psi|H|\psi\rangle}{{\langle \psi|\psi\rangle}}-E_{\text{HF}}~,
\eeq
relative to its exact value for $L=12$ at half filling. On the one hand, notice how PBCS (\ref{pbcs}) and CCD$_{\text{sep}}$ (\ref{sccd}) are limited in the weak pairing regime by their common separable structure, both reducing to $(1+\Gamma^\dagger_P\Gamma^\dagger_H)\hf$ at $G/\epsilon\ll1$. On the other hand, the full generality of the structure matrix of the vCCD wavefunction (\ref{ccd}) allows it to capture precisely all the correlations in this regime. However, beyond the critical value $G_{\text{cr}}$ of the HF to BCS transition the vCCD ansatz quickly loses its ability to describe the stronger correlations, becoming indistinguishable from its separable version  vCCD$_{\text{sep}}$ for $G/\epsilon >1$ (see also the discussion around Fig. 4 of Ref. \cite{duk16}).

As indicated by the thick green line next to the horizontal axis of Fig. (\ref{fig1}), the combined vCCD$_{\text{sep}}$PBCS wavefunction (\ref{sccdpbcs}) leads to very accurate energetics across all regimes with the relative errors in the correlation energy not exceeding $3\cdot 10^{-4}$. Note however that it is essential to enable all parameters to vary freely in the minimization process as to retain the full flexibility of the wavefunction and thus recover all available dynamical correlations. Indeed, by considering a variational CCD$_{\text{sep}}$ on top of the frozen optimal PBCS reference (or viceversa) we find significant improvements only in the weak pairing regime, as indicated by the dashed lines in Fig. (\ref{fig1}). In particular, by varying in this way just the PBCS amplitudes within vCCD$_{\text{sep}}$PBCS while keeping fixed the optimal CCD$_{\text{sep}}$ structure (computed beforehand) we find no improvement over the full vCCD results for $G>G_{\text{cr}}$. At $G/\epsilon\ll1$, the results differ due to the limitations discussed around Eq. (\ref{sccdpbcs_weak}).

\begin{figure}[t]
\centering
\includegraphics[width=0.5\textwidth]{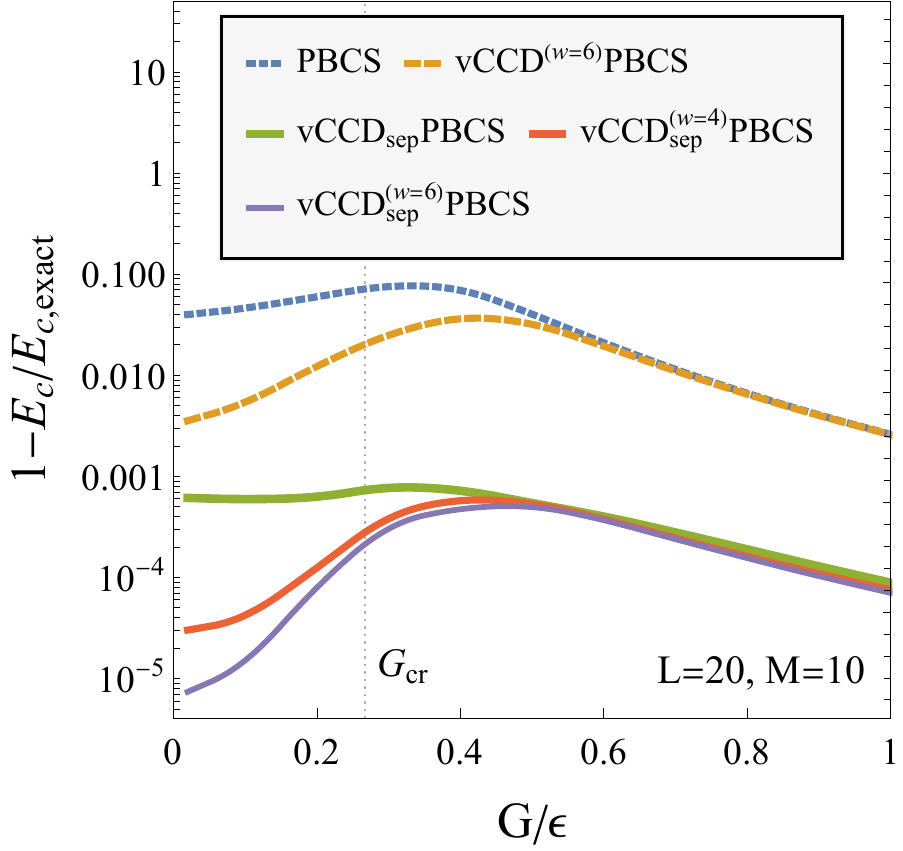}
\caption{Error in the correlation energy  (\ref{corr_energy}) relative to its exact value (in percentages) versus the pairing strength $G$ (in units of the level spacing $\epsilon$) for the fully variational PBCS (\ref{pbcs}), vCCD$_{\text{sep}}$PBCS (\ref{sccdpbcs}),  vCCD$^{(w=6)}$PBCS (\ref{ccdwpbcs}) and vCCD$^{(w=4,6)}_{\text{sep}}$PBCS (\ref{ccdwseppbcs}) wavefunctions in the case of an $L=20$ level system at half filling.}
\label{fig2}
\end{figure}

\begin{figure*}[t]
\centering
\includegraphics[width=\textwidth]{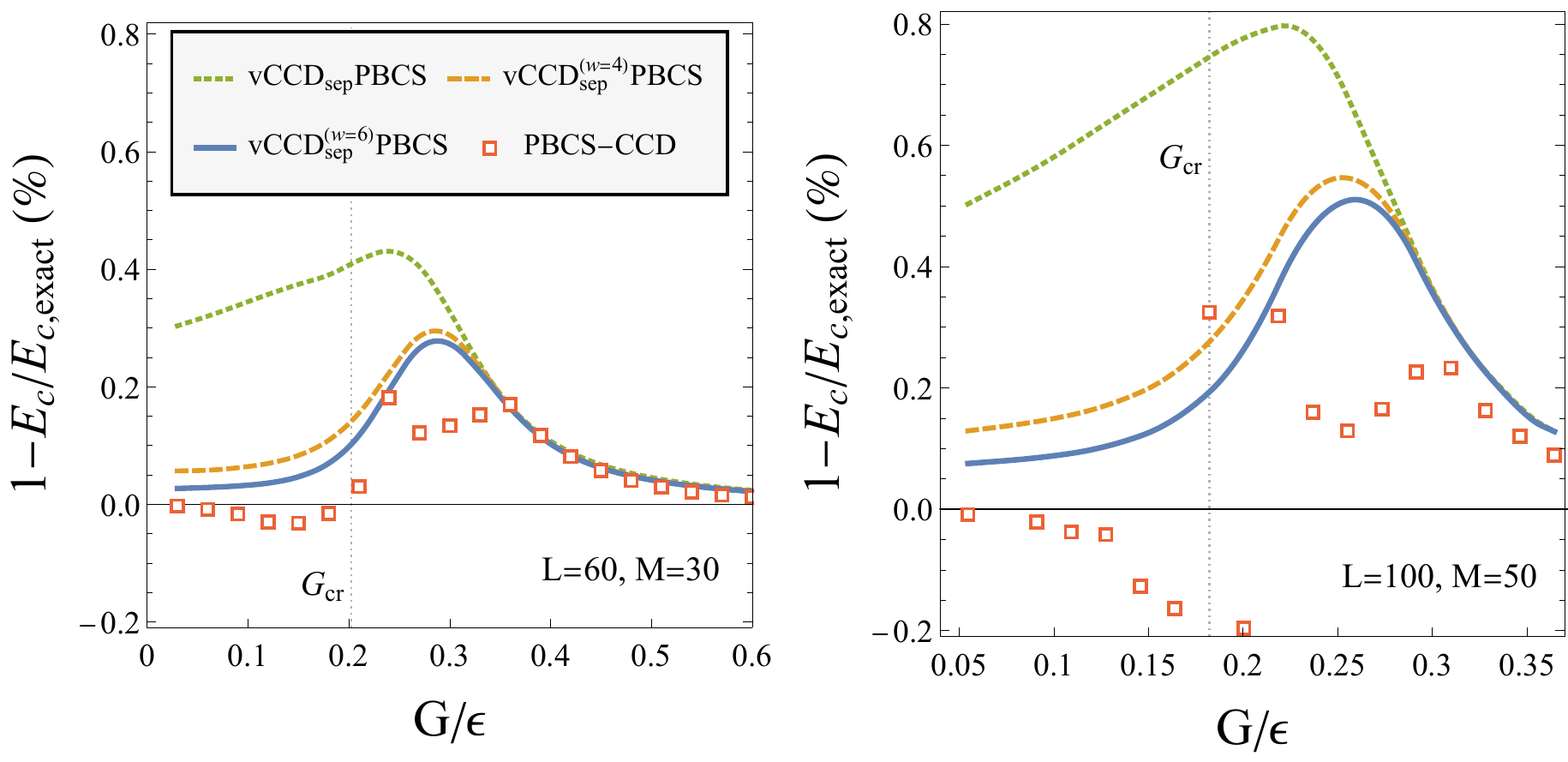}
\caption{Error in the correlation energy  (\ref{corr_energy}) relative to its exact value (in percentages) versus the pairing strength $G$ (in units of the level spacing $\epsilon$) for the fully variational vCCD$_{\text{sep}}$PBCS (\ref{sccdpbcs}) and vCCD$^{(w)}_{\text{sep}}$PBCS (\ref{ccdwseppbcs}) wavefunctions, in the case of an $L=60$ level system (left panel) and for an $L=100$ level system (right panel) at half filling. The red squares indicate the results of the particle-number projected Bogoliubov-coupled-cluster theory of Ref. \cite{qiu2019}. Note also that for $L=100$, the errors for the pure PBCS (\ref{pbcs}) reach a maximum of about 22\% at $G/\epsilon=0.24$ and decrease down to 6\% at very small $G$ (see also Fig. 1 of Ref. \cite{qiu2019}).}
\label{fig3}
\end{figure*}

The relative errors for the correlation energy (\ref{corr_energy}) relative to its exact value are shown in logarithmic scale in Fig. (\ref{fig2}) for a slightly larger system of $L=20$ levels at half-filling. Note that the vCCD$_{\text{sep}}$PBCS errors have slighly increased with respect to the $L=12$ case, but are still respectably accurate when compared to PBCS. The limitations of the vCCD$_{\text{sep}}$PBCS wavefunction in the weak pairing regime originate in the approximate form of its structure matrix, as remarked around Eq. (\ref{sccdpbcs_weak}).

This weak pairing behaviour may be  improved at a reasonable computational cost by including additional non-separable CCD excitations on top of vCCD$_{\text{sep}}$PBCS. The vCCD$^{(w)}_{\text{sep}}$PBCS ansatz (\ref{ccdwseppbcs}) already provides more than an order of magnitude lower errors with respect to vCCD$_{\text{sep}}$PBCS at weak pairing  even for the modest $w=4$. A larger non-separable CCD excitation window $w=6$  naturally accounts for an additional amount of correlations, further reducing the errors at weak pairing. Beyond the critical value $G_{\text{cr}}$ however there is no significant benefit of the supplementary non-separable excitations. This is also the situation for the vCCD$^{(w)}$PBCS ansatz (\ref{ccdwpbcs}) which displays the same strong pairing behaviour as the plain PBCS. With the Fermi sea being washed out at strong pairing, only a global deformation of PBCS such as vCCD$_{\text{sep}}$PBCS is able to bring substantial improvement to the energetics, as opposed to any local deformation such as vCCD$^{(w)}$PBCS (\ref{ccdwpbcs}).

The numerical values for the optimal non-separable structure matrix elements $z_{ph}$ are typically found to be very small, of the order $10^{-2}-10^{-3}$ upon the full vCCD$^{(w)}_{\text{sep}}$PBCS energy minimization. In practice, within the vCCD$^{(w)}_{\text{sep}}$PBCS approach we chose to limit the non-separable CCD excitations to linear order
\beq
\label{linear_ccdw}
\begin{aligned}
&\text{vCCD}^{(w)}(z)=1+\sum_{p=M+1}^{M+w/2}\sum_{h=M+1-w/2}^Mz_{ph} P^\dagger_p P^\dagger_h~,
\end{aligned}
\eeq
which allows for a decrease in the computational complexity without any noticeable loss in precision with respect with the full form of Eq. (\ref{ccdwpbcs}).

An additional computational speed-up is enabled by limiting the action of the particle number projection operations in Eqs. (\ref{pbcs_expans}) and (\ref{Gamma_ccd_expans}) to a reduced subspace of particle and hole collective pairs. For the moderate values  of the pairing strength $G$ considered here (2 to 3 times the value of $G_{cr}$) this still allows for an exact particle-number conservation (within the numerical accuracy). This key computational aspect is ensured by the negligible contributions of the high order terms within the PBCS (\ref{pbcs}) and vCCD$_{\text{sep}}$ (\ref{sccd_op}) expansions, due to the small (subunitary) numerical values of their corresponding collective pair amplitudes. All results presented below for $L=60$ and $L=100$ were obtained upon projecting the particle-number only within a $\mathcal{C}_1=15$ ph-pair subspace for PBCS in Eq.  (\ref{pbcs_expans}) and within a $\mathcal{C}_2=10$ ph-pair subspace for CCD$_{\text{sep}}$  in Eq. (\ref{Gamma_ccd_expans}). The gauge angles were adjusted accordingly, i.e. $\theta_n=2\pi n/(\mathcal{C}_1+1)$, $\phi_k=\varphi_k=2\pi k/(\mathcal{C}_2+1)$. We also note that for all considered systems it was sufficient to restrict the amount of CCD$_{\text{sep}}$ excitations by imposing an upper bound at $\ell_{\text{max}}=7$ in the $\ell$-sum of Eq. (\ref{expans_coeffs_sccd}), as to avoid the numerical errors originating from the combination of large factorials and rapidly oscillating phases and, at the same time, but preserving the numerical precision.

We present in Fig. (\ref{fig3}) the relative errors in the correlation energy (\ref{corr_energy}) relative to its exact values for $L=60$ (left panel) and $L=100$ (right panel) at half filling. The vCCD$_{\text{sep}}$PBCS wavefunction remains surprisingly accurate given that its doubly-separable structure matrix (\ref{sccdpbcs_weak}) at weak coupling involves only $L$ independent parameters {(down from a total of $2L$ parameters due to the present ph-symmetry)} as compared to the fully non-separable CCD structure matrix that requires $L^2/8$ independent parameters (still for the present ph-symmetric systems).  More precisely, the  vCCD$_{\text{sep}}$PBCS errors at weak coupling are 0.3\% and 0.5\% for $L=60$ and $L=100$ respectively, increasing slightly until $G\sim G_{\text{cr}}$ and then rapidly decreasing at stronger couplings.

While the inclusion of non-separable local CCD excitations within the vCCD$^{(w)}_{\text{sep}}$PBCS ansatz (\ref{ccdwseppbcs})  significantly improves the error at weak coupling even for relatively very small values of $w$, perfectly accurate energetics would still require a set of global ph-excitations on top of vCCD$_{\text{sep}}$PBCS. Nevertheless, the overall quality of the vCCD$^{(w=6)}_{\text{sep}}$PBCS results is at the level of the more involved particle-number projected Bogoliubov-coupled-cluster theory of Ref. \cite{qiu2019}, while consistently providing an energy upper bound across all regimes.

\begin{figure}[t]
\centering
\includegraphics[width=0.5\textwidth]{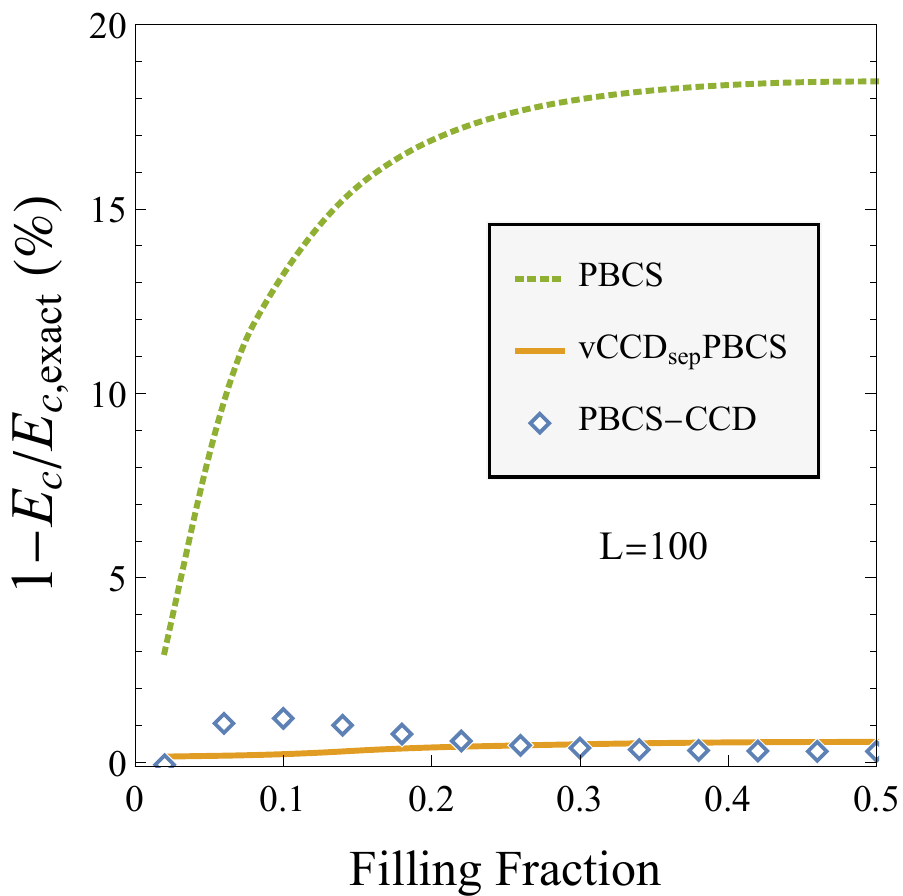}
\caption{Error in the correlation energy  (\ref{corr_energy}) relative to its exact value (in percentages) versus the filling fraction $M/L$ of an $L=100$ level system for the fully variational PBCS (\ref{pbcs}) and vCCD$_{\text{sep}}$PBCS (\ref{sccdpbcs}) wavefunctions. The pairing strength is $G=1.5G_{\text{cr}}$, with $G_{\text{cr}}$ the critical strength at half filling. The blue diamonds indicate the results of the particle-number projected Bogoliubov-coupled-cluster theory of Ref. \cite{qiu2019}.}
\label{fig4}
\end{figure}

The situation is similar away from half-filling, as shown in Fig. (\ref{fig4}). To provide a comparison with the results of Ref. \cite{qiu2019}, we consider the $L=100$ case at an interaction strength of $G=1.5 G_{\text{cr}}$. For this value the particle-number symmetry is broken for all filling fractions. Both PBCS and vCCD$_{\text{sep}}$PBCS are most accurate for small (or large) filling fractions, but their errors live on different scales (0.5\% vs 20\% at half filling). As seen in Fig. (\ref{fig3}), in the considered regime the non-separable local CCD excitations bring no improvement over the vCCD$_{\text{sep}}$PBCS results, which are well matched against those of Ref. \cite{qiu2019}.

Finally, we consider the behaviour of the canonical gap
\beq
\label{gap}
\Delta=G\sum_{i=1}^L \sqrt{n_i(1-n_i)}~,
\eeq
where $n_i=\langle c^\dagger_i c_i + c^\dagger_{\bar{i}} c_{\bar{i}}\rangle/2$ indicates the occupation probability of each level $i$. This quantity exhibits a more pronounced sensitivity to the structure of the wavefunction than the correlation energy due to its dependence on the occupation probabilities.

\begin{figure}[ht!]
\centering
\includegraphics[width=0.5\textwidth]{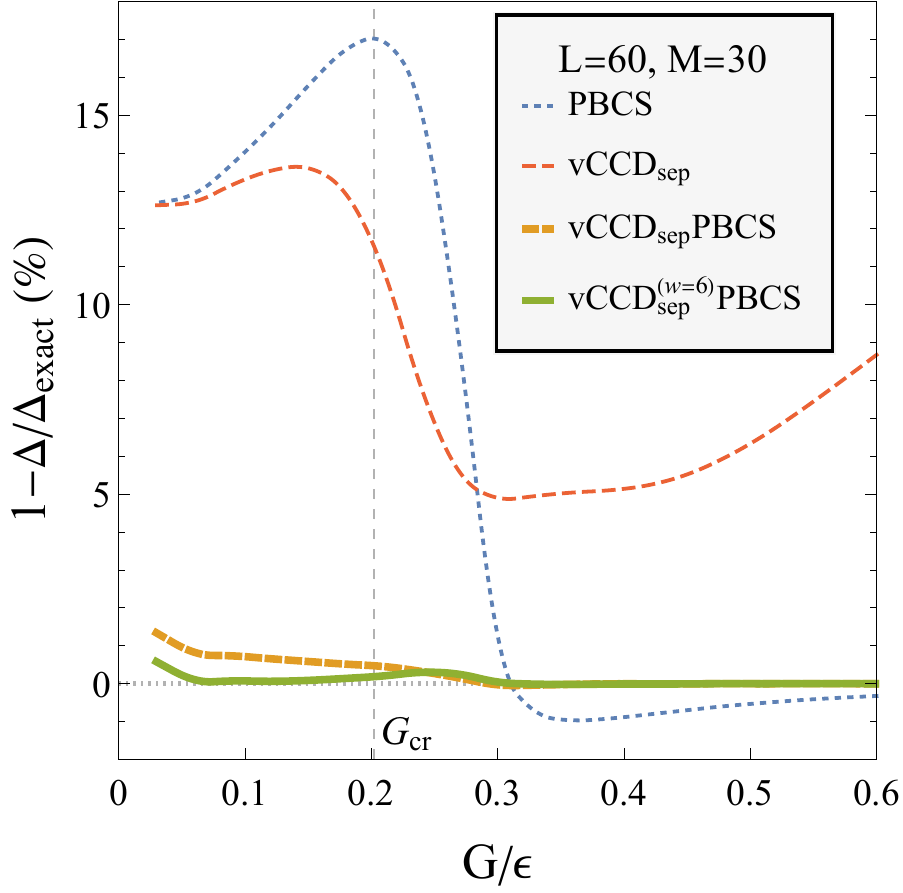}
\caption{Error in the canonical gap (\ref{gap}) relative to its exact value (in percentages) versus the pairing strength $G$ (in units of the level spacing $\epsilon$) for the fully variational PBCS (\ref{pbcs}), vCCD$_{\text{sep}}$ (\ref{sccd}),  vCCD$_{\text{sep}}$PBCS (\ref{sccdpbcs}) and vCCD$^{(w=6)}_{\text{sep}}$PBCS (\ref{ccdwseppbcs}) wavefunctions in the case of an $L=60$ level system at half filling.}
\label{fig5}
\end{figure}

We show in Fig. (\ref{fig5}) the error for the canonical gap $\Delta$ relative to its exact value for $L=60$ at half filling. Improving on both individual PBCS and vCCD$_{\text{sep}}$ wavefunctions, their combination vCCD$_{\text{sep}}$PBCS only shows visible $\sim1\%$ errors in the $G<G_{\text{cr}}$ region. These are further reduced within the vCCD$^{(w)}_{\text{sep}}$PBCS approach which exhibits highly accurate occupations across all regimes.

\section{SUMMARY AND CONCLUDING REMARKS
}

In this work, we considered a variational approach for the ground state of finite paired systems based on a combined coupled-cluster and particle-number-projected BCS wavefunction. We benchmarked our results against the exact solution for a picket-fence model involving a pure pairing force acting on a space of doubly
degenerate, equally distanced levels. The analyzed systems systems range from small ($L=M=12$) to relatively large ($L=M=100$).

We confirmed within the variational context that the combination of symmetry-restored mean field theory and coupled-cluster
theory leads to a wavefunction that is significantly better than either of the two taken separately, which is also the main conclusion of Ref. \cite{qiu2019}. By incorporating pure four-body correlations,  our vCCD$^{(w)}_{\text{sep}}$PBCS (\ref{ccdwseppbcs}) is able to reproduce well the physics at weak pairing while also offering orders of magnitude smaller errors relative to the standard pair-only PBCS (\ref{pbcs}) across all other regimes. Our results match the high level of precision of the particle-number projected Bogoliubov-coupled-cluster theory of Ref. \cite{qiu2019}, while consistently providing an energy upper bound.

Computational limitations include the restriction of the pure quartet correlations to a relatively small window around the Fermi level (discussed in detail in the main text) and also the need of a restricted space for particle-number projection. While the latter does not spoil the exact particle-number conservation (within the numerical accuracy) for moderate values of the pairing strength, alternative approaches need to be considered for the very strong pairing regime $G \gg G_{\text{cr}}$. One possibility would involve limiting the separable CCD excitations to linear order within the vCCD$_{\text{sep}}$PBCS approach. Computations with this simpler wavefunction could then be performed efficiently without resorting to any approximations for an improvement over the already very good PBCS results for this regime.

The  vCCD$^{(w)}_{\text{sep}}$PBCS (\ref{ccdwseppbcs}) wavefunction was shown to be quite flexible but it is still affected by structural limitations leading to a visible maximum in the energy error around $G\sim1.5G_{\text{cr}}$ in all analyzed cases. Attempts at mitigating these effects have included treating as independent variational parameters the factorials in Eq. (\ref{expans_coeffs_sccd}) originating from the vCCD$_{\text{sep}}$ expansion (\ref{sccd}), with only marginal benefits.

Possible avenues to explore could involve the variational treatment of beyond-ph CCD excitations on top of the symmetry-restored mean-field state as in
\beq
|\Psi\rangle=\exp\left(\sum_{i,j=1}^L z_{ij}\,c^\dagger_{i}c^\dagger_{\bar{i}}\, c_{\bar{j}} c_j\right) \pbcs~,
\eeq
which would act as a more natural choice (albeit more computationally challenging) for a regime lacking a well-defined Fermi sea. As a first step one could envision building a multi-separable beyond-ph CCD excitation operator which could be optimized with the very recently developed methods for constructing linearly-independent PBCS states \cite{dutta2021}.

Finally, we leave to future studies extensions of the theory presented in this work that could incorporate the effect of seniority breaking terms within a generalized variational wavefunction.

\begin{acknowledgments}
This work was supported by a grant of the Romanian Ministry of Education and Research, CNCS - UEFISCDI,
project number PN-III-P1-1.1-PD-2019-0346, within PNCDI III, and PN-19060101/2019-2022; and by the Spanish Ministerio de Ciencia e Innovaci\'on, and the European regional development fund (FEDER) project Nº PGC2018-094180-B-I00. We thank the authors of Ref. \cite{qiu2019} for sharing with us their numerical results. 
\end{acknowledgments}

\bibliography{mybibfile_v4}

\end{document}